# Student's readiness for e-learning in the Universities in Yemen

**Adnan Sharaf Ali Yousef Al-Absi[1], Ivelina Peneva[2], Krasimir Yordzhev[3]**

*South-West University "Neofit Rilski", Blagoevgrad, Bulgaria*
[1] engadnansharaf@yahoo.com, [2] ivelina_peneva@abv.bg, [3] yordzhev@swu.bg

***Abstract***: *The e-learning is an advanced version of the traditional education. It's defined as a way of learning by using the communication mechanisms of modern computer networks and multimedia, including voice, image, and graphics and mechanisms to search electronic libraries, as well as web portals, whether in the context of distance learning or in the classroom. The people engage in the transition to web-supported education are the administrative staff, the faculty, and the students. They all have their needs and they all should meet specific requirements in order to facilitate the transition. The article presents the results of questionnaire research of the student's readiness for e-learning in Yemeni universities.*

## Introduction

A survey conducted with 230 students at Sana University and at the Technology and Science University of Yemen, makes it clear that, as a whole, learners have a positive evaluation of the integration IT in education. The choice of informants has been dictated by the fact that the students at these two universities come from a variety of ethnic, cultural, and religious backgrounds. In addition, these universities seem to host the most students and many of them are women, which is not common to all Yemeni universities.

## Methodology

The study was conducted through modification of a questionnaire to study students' readiness for mobile learning [2]. To ensure the reliability property, the reliability coefficients for each axis of the study axes have been extracted by using Alpha Cronbach [4]. It is observed that the reliability coefficients is not less than the minimum acceptable to judge the extent to which the tool has reliability property (0.6), thus the reliability property is present. Therefore, the tool is valid, suitable for field application, and measuring.

## Results and discussion

The following table presents a demographic analysis of the participants in terms of their gender, study level, and field of studies (Table 1).

**Table 1:**

| | | Frequency | Percent % |
|---|---|---|---|
| study level | third -Year | 129 | 56.1 |
| | fourth year | 101 | 43.9 |
| field of study | Applied | 95 | 41.3 |
| | humanitarian | 135 | 58.7 |
| gander | female | 87 | 37.8 |
| | Male | 143 | 62.2 |

## What is the students' evaluation of the infrastructure (computer labs - the Internet - the library) in the faculties of Yemeni universities?

To answer this question, it has been extracted the arithmetic mean and standard deviation of the degree of participants' response to the level of the total degree and of the statements' axis of evaluation as can be seen from the Table 2.

**Table 2:**

| \multicolumn{5}{c}{**Axis I: Evaluation of the students of the infrastructure** (computer labs - the Internet - the library) in the faculties of Yemeni universities} |
|---|---|---|---|---|
| Order | Statement | Mean | Std. Deviation | Verbal meaning |
| 1A | A lab specialist is available to assist students. | 2.58 | .569 | Yes |
| A2 | Adequate lab space for students. | 2.56 | .531 | Yes |
| A3 | Computers in the computer lab at the college linked to an internal network. | 2.51 | .611 | Yes |
| A4 | Electric current is constantly available. | 2.51 | .527 | Yes |
| A5 | Internet service is available at the university library. | 2.50 | .543 | Yes |
| A6 | The number of computers in the lab on the number of students. | 2.47 | .581 | Yes |
| A7 | The computer lab is available in at the college Internet service. | 2.44 | .579 | Yes |
| A8 | Computers in the labs work well. | 2.40 | .596 | Yes |
| A9 | There is maintenance for equipment and accessories. | 2.38 | .668 | Yes |
| A10 | Programmes are new. | 2.28 | .615 | No |
| A11 | Library at your university has a sufficient scientific reference. | 2.27 | .643 | No |
| A12 | Computer lab in your college is linked to the main network of university. | 2.16 | .658 | No |
| A13 | There is updated for scientific references in the library. | 2.13 | .643 | No |
| A14 | Accessory devices like printers & scanners are available. | 2.13 | .414 | No |
| A15 | Library has is subscribed to scientific sites. | 2.12 | .643 | No |
| A16 | The speed of the internet is fast. | 2.08 | .474 | No |
| A17 | There are headsets in the labs. | 2.04 | .371 | No |
| A18 | Computer accessories like printers work well | 2.02 | .498 | No |
| A19 | There are additional facilities available for students with disabilities. | 2.00 | .443 | No |
| A20 | Printers can be used for research purposes. | 1.99 | .524 | No |

As can be seen from Table 2 the average of the total degree of the participants' response to the axis of the evaluation is (2.28), which is a value that indicates the second alternative answer (No). Therefore, it can be concluded that there is a semi full absence of infrastructure, which is necessary for the application of information technology in the Yemeni universities in general. At the level of all the statements in this axis, they have the arithmetic mean values ranging from (1.99) to (2.58). Some of these values refer to the first alternative (Yes) which is related to the statements that occupy the order (A1- A9), which is a positive answer and means that the application of information technology requirements is available in Yemeni universities that have been identified regarding the students' evaluation of the infrastructure. While we find that, the average values for the rest of the statements refer to the second alternative (No) which is related to the statements that occupy order (A10-A20). This is a negative answer and means that the application of information technology requirements is not available in Yemeni universities that have been identified regarding the students' evaluation of the infrastructure.

### Does the student have the necessary capacities for the use of information technology?

As can be seen from Table 3, the average of the total degree of the participants' response to the axis of the abilities of students in the use of information and communication technology is (2.50). It is a value that indicates the first alternative answer (Yes), which means that the students have the necessary abilities in the use of information and communication technology in general. At the level of all the phrases in these axis, they have the arithmetic mean values ranged from (2.14) and (2.79), most of these values refer to the first alternative (Yes) which is related to statements (B1- B7), where is a positive answer means that the students have the abilities in the use of information and communication technology in Yemeni universities. On the other hand, we find that the average values for the two statements which come last in order refer to the second alternative (No), where a negative answer means that the students do not have the abilities in the use of information and communication technology. These have been identified as (B8 and B9)**.**

**Table 3:**

| Axis II: the abilities of students in the use of information and communication technology ||||| 
|---|---|---|---|---|
| Order | Statement | Mean | Std. Deviation | Verbal meaning |
| B1 | Using word processor to prepare a CV or a type a research report etc. | 2.79 | .431 | Yes |
| B2 | Using Email to send attachments | 2.75 | .452 | Yes |
| B3 | Using PowerPoint for presentations | 2.74 | .503 | Yes |
| B4 | Using the Internet to find digital database. | 2.55 | .587 | Yes |
| B5 | Using e-mail to communicate with the teacher by the Internet. | 2.50 | .582 | Yes |
| B6 | Using Discussion Forums to learn online. | 2.43 | .555 | Yes |
| B7 | Using an interactive site for learning the Internet. | 2.43 | .593 | Yes |
| B8 | Using video conferencing for online learning. | 2.16 | .516 | No |
| B9 | Using Curriculum is based on the Web. | 2.14 | .484 | No |
| Students' abilities in the use of information and communication technology || 2.50 | .327 | Yes |

### What is the perception of students towards the education based on information technology as compared with traditional education?

As can be seen from the Table 4, the average of the total degree of the participants' response to the axis of the perception of students towards education based on information technology compared with traditional education was (3.87), which is a value that indicates the second alternative answer (Agree). This means that the perception of students towards education based on information technology compared with traditional education is positive in general.

**Table 4:**

|  | **Axis III: Perceptions of students about education base on Information and communication technology compared to the traditional education** |  |  |  |
|---|---|---|---|---|
| Order | Statement | Mean | Std. Deviation | Verbal meaning |
| 1C | I believe that the information and communication technologies contribute to the improvement of education. | 4.51 | .685 | Strongly Agree |
| C2 | I think that the video and audio texts can improve my level of education. | 4.46 | .823 | Strongly Agree |
| C3 | Learning based on ICT helps the exchange of information between universities. | 4.35 | .661 | Strongly Agree |
| C4 | I would like to study by the computer even if it may be difficult. | 4.32 | .742 | Strongly Agree |
| C5 | Information and communication technology allows exchanging information in an effective manner. | 4.30 | .863 | Strongly Agree |
| C6 | Expanding the scope of resources and exchange information available to students by using the internet. | 4.19 | .866 | Agree |
| C7 | A media of information technology provides an opportunity for students in remote areas who could not get a regular education to continue their studies. | 4.17 | .785 | Agree |
| C8 | A media of information technology will help students to identify the institutions of higher education in their own countries and abroad | 4.17 | .789 | Agree |
| C9 | Learning by online helps to exchange information between the student and the teacher. | 4.12 | .883 | Agree |
| C10 | Learning by ICT requires high computer skills. | 3.80 | 1.059 | Agree |
| C11 | Learning by computer lacks the interaction between the student and the teacher. | 3.72 | 1.127 | Agree |
| C12 | It is difficult to find good quality information on the internet. | 2.93 | 1.169 | Undecided |
| C13 | Learning through information and communication technology takes more time than the traditional method. | 2.61 | 1.135 | Undecided |
| C14 | I prefer learning by the traditional method, i.e. from books and not computers | 2.53 | 1.076 | Disagree |

At the level of all the statements in this axis, the arithmetic mean values range from (2.53) and (4.51). Some of these values refer to the first alternative (Strongly Agree) which refers to the statement that occupy the order (C1- C5), where a positive answer means that the students have high positive perception towards the education based on information technology compared with traditional education in Yemeni universities which. These have been identified as follows (C1-C5). On the other hand, we find that the average values for the rest of the statements refer to the second alternative (Agree) which are statements (C6-C11), where a negative answer means that the students have positive perception to some extent towards education based on information technology compared with traditional education in the Yemeni universities.

Additionally, it should be noted that the average values of statements (C12) and (C13) before the final respectively indicate to the third alternative (Undecided), where the answer Undecided means that the perceptions of students does not show a clear tendency towards the education-based information and communication technology compared to traditional education found in (C12 and C13).

In addition, the last statement refers to the value of its arithmetic mean to the fourth alternative (Not-Agree), where a negative answer means that the students have negative perception of education based on information and communication technology compared to traditional education as shown in (C14).

**Test hypotheses.**

**First hypothesis: There are not statistically significant differences in the respondents' response to the study axes due to the variable academic level of the student.**

To ensure the correctness of this hypothesis was used Two Independent Samples T-Test [5].

**Table 5:**

| AXIS | Study level | N | Mean | Std. Deviation | T | df | Sig. (2-tailed) |
|---|---|---|---|---|---|---|---|
| Evaluation of the students of the infrastructure | third year | 129 | 2.28 | .280 | .296 | 228 | .767 |
| | fourth year | 101 | 2.27 | .303 | | | |
| The abilities of students in the use of information and communication technology | third year | 129 | 2.44 | .343 | 2.962 | 228 | .003 |
| | fourth year | 101 | 2.57 | .290 | | | |
| Perceptions about students based education Information and communication technology compared to the traditional education | third year | 129 | 3.84 | .409 | 1.046 | 228 | .297 |
| | fourth year | 101 | 3.90 | .411 | | | |

The test's results indicated in the Table 5 show that the level of significance values (Sig) is less than (0.05), at the level of the axis of students' abilities in the use of information and communications technology which means that there are statistically significant differences in students' abilities in the use of information and communication technology due to the variable academic level. These differences are in favour of students in the fourth academic year as can be seen from the value of the arithmetic average of this category, which is larger than the value of their peers from the third level.

While it is observed that the value of the level of significance at the level of the axis of students' evaluation of the infrastructure and the axis of the perceptions of students about education based on information and communication technology compared to the traditional education was greater than (0.05), it means that there are not statistically significant differences at the level of axes due to the variable the academic level of the student.

**The second hypothesis: There are not statistically significant differences in the participants' response to the study axes due to the variable academic specialization for the student.**

To ensure the correctness of this hypothesis was used Two Independent Samples T-Test [5]. The results can be seen in Table 6.

**Table 6:**

| AXIS | field of study | N | Mean | Std. Deviation | T | df | Sig. (2-tailed) |
|---|---|---|---|---|---|---|---|
| Evaluation of the students of the infrastructure | Applied | 95 | 2.31 | .247 | 1.798 | 228 | .074 |
| | Humanitarian | 135 | 2.25 | .314 | | | |
| The abilities of students in the use of information and communication technology | Applied | 95 | 2.61 | .284 | 4.737 | 228 | .000 |
| | Humanitarian | 135 | 2.41 | .330 | | | |
| Perceptions of students about education base on Information and communication technology compared to the traditional education | Applied | 95 | 3.97 | .364 | 3.395 | 228 | .001 |
| | Humanitarian | 135 | 3.79 | .424 | | | |

The test's results indicated in the Table 6 show that the level of significance values (Sig) is less than (0.05), at the level of the axis of students' abilities in the use of information and communications technology and the axis of the perceptions of students about education based on information and communication technology compared to the traditional education, which means that there are statistically significant differences in participants' response in the two axes due to the variable the academic specialization, these differences in favour of students in the specialization (Applied) can be seen from the value of the arithmetic average of this category, which is larger than the value of their peers in the specialty (humanitarian).

Importantly, it is observed that the value of the level of significance at the level of the axis of students' evaluation of the infrastructure was greater than (0.05), it means that there are not statistically significant differences at the level of axis due to the variable academic specialization.

**The third hypothesis: There are not statistically significant differences in the participants' response to the study axes due to the variable gender of the student?**

To ensure the correctness of this hypothesis was used Two Independent Samples T-Test [5]. The results can be seen in Table 7.

**Table 7:**

| AXIS | Gander | N | Mean | Std. Deviation | T | df | Sig. (2-tailed) |
|---|---|---|---|---|---|---|---|
| Evaluation of the students of the infrastructure | Female | 87 | 2.22 | .256 | 2.029 | 228 | .044 |
| | Male | 143 | 2.30 | .305 | | | |
| The abilities of students in the use of information and communication technology | Female | 87 | 2.41 | .324 | 3.182 | 228 | .002 |
| | Male | 143 | 2.55 | .318 | | | |
| Perceptions about students based education Information and communication technology compared to the traditional education | Female | 87 | 3.85 | .343 | .544 | 228 | .587 |
| | Male | 143 | 3.88 | .447 | | | |

The test's results indicated in the Table 7 show that the level of significance values (Sig) is less than (0.05) at the level of the axis of students' abilities in the use of information and communications technology and the axis of students' evaluation of the infrastructure, which means that there are statistically significant differences in respondents' response in the two axes due to the variable gender. These differences are in favour of (Male) as can be seen from the value of the arithmetic average of this category, which is larger than the value of their peers (female).

While it is observed that the value of the level of significance at the level of the axis of the students' perceptions about education based on information and communication technology compared to traditional education is greater than (0.05), meaning that there are not statistically significant differences at the level of axis due to the variable the gender.

**Summary**

The successful integration of information technology in higher education will contribute to the solution of many problems facing developing countries [3], H., 2005). Admittedly, there are many problems on the way, such as lack of investment in physical assets; scarcity of qualified academic staff; hesitation of some girls to register in universities due to the conservative culture society.

Still, the application of Information technology in higher education in Arab countries, and especially in Yemen, should not be based on technical decision but on strategic planning as a national choice to improve higher education so it can meet the economic and social development needs [1].